\RequirePackage{lineno}
\documentclass[aps,prl,twocolumn,superscriptaddress,groupedaddress,preprintnumbers]{revtex4}  
\usepackage{graphicx}  
\usepackage{dcolumn}   
\usepackage{bm}        
\usepackage{amssymb}   
\usepackage{multirow}
\usepackage{epsfig}
\usepackage{amsmath}
\usepackage[colorlinks, linkcolor=blue]{hyperref}
\usepackage{ulem}

\newcommand{\effstw}{\ensuremath{\sin^2\theta_{\text{eff}}^{\text{$\ell$}}}}

\begin{document}

\lefthyphenmin=2
\righthyphenmin=2

\widetext

\hfill {\large MSUHEP-22-031}

\title{ Measurement of the proton structure parameters in the forward-backward charge asymmetry}
\affiliation{Department of Modern Physics, University of Science and Technology of China, Jinzhai Road 96, Hefei, Anhui 230026, China}
\affiliation{School of Nuclear Science and Technology, University of South China, Hengyang, Hunan 421001, China}
\affiliation{School of Physics Science and Technology, Xinjiang University, Urumqi, Xinjiang 830046, China}
\affiliation{Department of Physics and Astronomy, Michigan State University, East Lansing, MI 48823, USA}

\author{Mingzhe Xie} \affiliation{Department of Modern Physics, University of Science and Technology of China, Jinzhai Road 96, Hefei, Anhui 230026, China}
\author{Siqi Yang}\email{yangsq@ustc.edu.cn} 
\affiliation{Department of Modern Physics, University of Science and Technology of China, Jinzhai Road 96, Hefei, Anhui 230026, China}
\author{Yao Fu} \affiliation{Department of Modern Physics, University of Science and Technology of China, Jinzhai Road 96, Hefei, Anhui 230026, China}
\author{Minghui Liu} \affiliation{Department of Modern Physics, University of Science and Technology of China, Jinzhai Road 96, Hefei, Anhui 230026, China}
\author{Liang Han} \affiliation{Department of Modern Physics, University of Science and Technology of China, Jinzhai Road 96, Hefei, Anhui 230026, China}
\author{Tie-Jiun Hou} \affiliation{School of Nuclear Science and Technology, University of South China, Hengyang, Hunan 421001, China}
\author{Sayipjamal Dulat} \affiliation{School of Physics Science and Technology, Xinjiang University, Urumqi, Xinjiang 830046, China}
\author{C.-P. Yuan} \affiliation{Department of Physics and Astronomy, Michigan State University, East Lansing, MI 48823, USA}

\begin{abstract}
The forward-backward asymmetry ($A_{FB}$) in the Drell-Yan process $pp/p\bar{p} \rightarrow Z\gamma^* \rightarrow \ell^+\ell^-$ is sensitive to 
the proton structure information. Such information has been factorized into well-defined proton structure parameters which can be 
regarded as experimental observables. In this paper, we extract the structure parameters from the $A_{FB}$ distributions reported by 
the CMS collaboration in $pp$ collisions at $\sqrt{s}=8$ TeV, and by the D0 collaboration in $p\bar{p}$ collisions at $\sqrt{s} = 1.96$ TeV. 
It is the first time that the unique parton information in the $A_{FB}$ spectrum can be decoupled from the electroweak calculation and measured 
as standalone observables, which can be used as new data constraints in the global quantum chromodynamics 
analysis of the parton distribution functions (PDFs). 
Although the parton information in the $pp$ and $p\bar{p}$ collisions are different, and the precisions of the measured structure parameters are 
statistically limited, the results from both the hadron colliders indicate that the down quark contribution might be higher than the 
theoretical predictions 
with the current PDFs at the relevant momentum fraction range.
\end{abstract}
\maketitle

\section{Introduction}

The forward-backward asymmetry ($A_{FB}$) of the Drell-Yan process $pp/p\bar{p}\rightarrow Z/\gamma^* \rightarrow \ell^+\ell^-$ 
is proved to be sensitive to the proton structure information, and could have important impact on the global 
quantum chromodynamics (QCD) analysis of the parton distribution 
functions (PDFs)~\cite{ArieStudy, ATLASStudy1, ATLASStudy2, CPCstudy}. Although $A_{FB}$ has been measured with quite a good 
precision at both the Tevatron and the Large Hardon Collider (LHC), the results are not yet included in the global QCD analysis 
of PDFs. The difficulty 
is the correlation between the proton structure information and the electroweak (EW) contribution in the $A_{FB}$ measurement, 
which 
causes large uncertainties extrapolating from one to the other~\cite{CPCstudy}. In the global analysis of NNPDF4.0~\cite{NNPDF4}, 
it is clearly stated that the $A_{FB}$ spectrum observed at the LHC has to be removed from the data set due to the difficulties in 
handling the correlation. 

In a recent study~\cite{AFBFactorization}, the proton structure information in the $A_{FB}$ spectrum has been factorized into 
well defined structure parameters, which 
can be used as new experimental observables and determined together with the effective weak mixing angle ($\effstw$), so that the 
correlation with the EW can be automatically taken into account. 

In this paper, we extract the structure parameters 
from the $A_{FB}$ distributions measured by the CMS collaboration using the $pp$ collision data at $\sqrt{s}=8$ TeV~\cite{CMSAFB8TeV}, 
and by the 
D0 collaboration using the $p\bar{p}$ collision data at $\sqrt{s}=1.96$ TeV~\cite{D0AFB5fb}. 
This work, of which the details will be discussed in the following sections, provides unique constraints on the proton 
structure information. Specifically, the structure parameters from the $A_{FB}$ separately reflect the contributions 
from both $u$ and $d$ type quarks, separately, 
which are always mixed in the total cross section measurements of the Drell-Yan production. 
As pointed out in Ref.~\cite{AFBFactorization} and Ref.~\cite{BoostAsymmetry}, these structure parameters can also constrain 
the dilution effect, which represents the contribution of a sea quark having higher energy than a valence quark in the initial state 
of the vector boson productions in $pp$ collisions. 

Although a complete global analysis of PDFs is needed to finally confirm the impact of the extracted structure parameters in this work,
the direct comparison between the measured values and their theoretical predictions already 
indicates that the down type quark contribution might be higher than the expectation at the relevant 
momentum fraction range, represented by the Bjorken variable $x$.
Such indication is consistent with the conclusion from the 
recent PDF global analysis that when the LHC data (other than $A_{FB}$) is included in the global fitting, 
the $d$ valence quark PDF becomes larger at $x$ around 0.1.~\cite{NNPDF4, CT18, MSHT20}. 
~\\

\section{Structure parameters obersvedd from the LHC data}

\begin{figure}[!hbt]
\begin{center}
\epsfig{scale=0.4, file=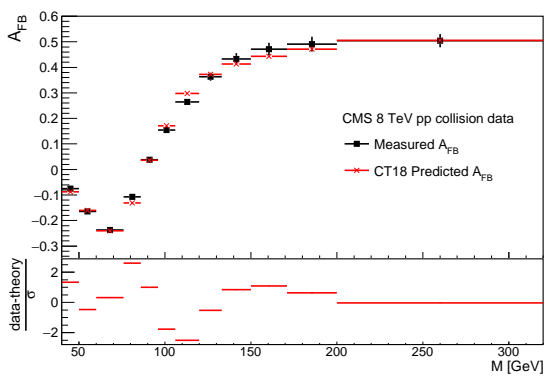}
\caption{\small The CMS 8 TeV $A_{FB}(M)$ measurement from the 
combined $e^+e^-$ and $\mu^+\mu^-$ 
events with $M=[40, 320]$ GeV and $1.5<|Y|<2.4$, compared to the {\sc ResBos} prediction 
with CT18NNLO PDFs. 
The bottom panel is the difference between the CMS measured 
$A_{FB}$ and the {\sc ResBos}+CT18 prediction, expressed in the unit of the 
total uncertainties $\sigma$, including the experimental uncertainty and the PDF uncertainty. 
The uncertainties on the theoretical predictions 
correspond to the $68\%$ C.L. PDF uncertainties.}
\label{fig:CMSAFBvsPDF}
\end{center}
\end{figure}

In this section, we discuss the extraction of the structure parameters using the $A_{FB}$ spectrum 
in $pp\rightarrow Z/\gamma^* 
\rightarrow \ell^+\ell^-$, ($\ell = e,\mu$) events measured by the CMS collaboration using the 
8 TeV $pp$ collision data~\cite{CMSAFB8TeV}. The $A_{FB}$ distributions 
are measured as a function of the dilepton mass ($M$) in a range of [40, 2000] GeV, 
and separately in five $Z$ boson rapidity bins ($Y$) of 
[0, 1], [1, 1.25], [1.25, 1.5], [1.5, 2.4] and [2.4, 5]. 
The central values and uncertainties of the observed asymmetry are provided, and the detailed numbers 
of the bin-by-bin correlations of systematics are given elsewhere~\cite{CMSMetaData}. 
The combined $A_{FB}$ of $e^+e^-$ and $\mu^+\mu^-$ events, and 
the corresponding uncertainties are replotted in Figure~\ref{fig:CMSAFBvsPDF}, together with the theoretical predictions 
in the rapidity bin $|Y|$ of [1.5, 2.4] as an example. 
The theoretical predictions are computed using the CT18 NNLO PDF~\cite{CT18}, 
and the {\sc ResBos}~\cite{resbos} package in which the QCD interaction is calculated at approximate next-to-next-to-leading order 
(NNLO) plus next-to-next-to-leading logarithm (NNLL), and the EW interaction is calculated based on the effective born approximation~\cite{PDG}, 
which gives precise predictions on the relationship between $A_{FB}$ and $\effstw$ around the $Z$ pole. 
The reported $A_{FB}$ distributions are unfolded to a phase space with no lepton acceptance cuts, thus 
the extracted structure parameters correspond to the same phase space in terms of $M$ and $Y$.

At the LHC, $A_{FB}$ is measured in the Collins-Soper frame~\cite{CSframe} with its $z$-axis 
defined according to the direction of the momentum of the dilepton system.
According to Ref.~\cite{AFBFactorization}, the observed $A_{FB}$ in a specific dilepton rapidity and mass configuration can be 
factorized as:
\begin{footnotesize}
\begin{eqnarray}\label{eq:LHCAFBfactorization}
A_{FB}(M) &=& \frac{\sum_{q=u,c} [1-2D_q(M)]\alpha_q(M)}{\alpha_\text{total}(M)} \cdot A^u_{FB}(M; \effstw) \nonumber \\
&+&  \frac{\sum_{q=d,s,b}[1-2D_q(M)]\alpha_q(M)}{\alpha_\text{total}(M)} \cdot A^d_{FB}(M; \effstw) \nonumber \\
&\equiv& [\Delta_u(M) + P^u_0] \cdot A^u_{FB}(M; \effstw) \nonumber \\
&+&  [\Delta_d(M) + P^d_0] \cdot A^d_{FB}(M; \effstw),
\end{eqnarray}
\end{footnotesize}
where $\alpha_q$ is the cross section of a specific subprocess with 
virtual photon and $Z$ boson coupled to $q\bar{q}$ $(q=u,d,s,c,b)$ in the initial state;
while $\alpha_\text{total}$ is the total cross section. 
$A^u_{FB}(M;\effstw)$ and $A^d_{FB}(M;\effstw)$ represent the original hard process asymmetries 
in the up-type and down-type subprocesses, respectively. Their calculations correspond to 
a special Collins-Soper frame, 
in which the directions of quark and antiquark are assumed to be known~\cite{AFBFactorization}. 
The values of 
$A^u_{FB}$ and $A^d_{FB}$ 
are solely determined by the single EW parameter of the effective weak mixing angle $\effstw$, which are independent 
of PDF. The dilution factors $D_q$ in the first equality of Eq.~\eqref{eq:LHCAFBfactorization}, is defined as 
the probability of having the antiquark energy higher than the quark energy, and can be modeled and predicted by the PDFs as: 
\begin{eqnarray}
D_q(x_L, x_S) &=& \frac{q(x_S)\bar{q}(x_L)}{ q(x_S)\bar{q}(x_L)+q(x_L)\bar{q}(x_S) }
\end{eqnarray}
where $x_S$ and $x_L$ are the Bjorken variables,  
respectively for the small and large values in a $q\bar{q}$ pair. 
They are related to the boson kinematics as $x_{L, S} = \sqrt{M^2 + Q^2_T}/\sqrt{s} 
\times e^{\pm Y}$, where $Q_T$ is the transverse momentum of the dilepton system. In Drell-Yan productions at the LHC, the larger 
fraction $x_L$ varies from $\mathcal{O}(10^{-2})$ to $\mathcal{O}(10^{-1})$, while the smaller one $x_S$ is at an order of 
$\mathcal{O}(10^{-4})$ to $\mathcal{O}(10^{-3})$. 

Based on the factorization formalism, the PDF information, that is presented by the cross sections $\alpha_q$ and dilution factors $D_q$, 
is thereafter decoupled from the EW calculations, as the coefficients in front of 
the $A^u_{FB}$ and $A^d_{FB}$ terms. 
The proton structure information can be further factorized as the structure parameters of $P^u_0$ and $P^d_0$ which are defined 
as the magnitude of the up-type and down-type coefficients, 
averaged over the mass range of the $A_{FB}$ spectrum, 
together with the residual mass-dependent 
terms of $\Delta_u(M)$ and $\Delta_d(M)$, respectively, as shown in the second equality 
of Eq.~\ref{eq:LHCAFBfactorization}. The detailed definition can be found in 
Ref.~\cite{AFBFactorization}. Since the dilution factors of the $s$, $c$ and $b$ quarks are close 
to 0.5, $P^u_0$ and $P^d_0$ are dominated by the $u$ and $d$ (anti)quark contributions.
In practice, the structure parameters $P^u_0$, $P^d_0$ and the EW parameter $\effstw$ can 
be treated as experimental observables, and determined by simultaneous fit to 
achieve the best agreement between the theoretical template of 
Eq.~\eqref{eq:LHCAFBfactorization} and the measured $A_{FB}(M)$ spectrum. Due to lack of sufficient constraints from the  
$A_{FB}$ distribution, the mass evolution terms $\Delta(M)$ have to be fixed to some PDF predictions. The PDF choice in 
the $\Delta(M)$ 
prediction would introduce additional theoretical uncertainties to the measurement of $P^u_0$, $P^d_0$ and $\effstw$ parameters. 
However, $\Delta(M)$ terms only describe the variation of parton densities in a relatively small mass window around 
the $Z$ pole
under investigation,  
thus the $\Delta$-induced uncertainties are not comparable to the statistical uncertainties 
of the data studied in this work.

\begin{figure}[!hbt]
\begin{center}
\epsfig{scale=0.4, file=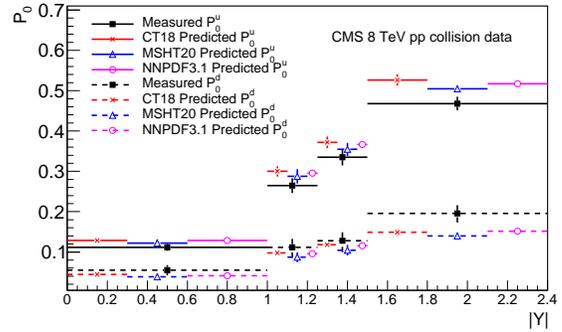}
\caption{\small The $P^u_0$ and $P^d_0$ parameters extracted from the $A_{FB}(M)$ spectrum 
in $|Y|$ bins of [0, 1.0], [1.0, 1.25], [1.25, 1.5] and [1.5, 2.4], and the corresponding 
{\sc ResBos} predictions from CT18, MSHT20 and 
NNPDF3.1. The error bars of the extracted $P^u_0$ and $P^d_0$ correspond to the uncertainty extrapolated from the total uncertainty 
and the bin-by-bin correlation provided by the CMS collaboration. The error bars of the predicted $P^u_0$ and $P^d_0$ 
correspond to the $68\%$ C.L. PDF uncertainties.}
\label{fig:CMSP0}
\end{center}
\end{figure}

Following the above strategy, the proton structure and EW parameters are then extracted from the CMS 
$e^+e^- + \mu^+\mu^-$ combined 8 TeV $A_{FB}$ 
~\cite{CMSAFB8TeV}, with four $|Y|$ bins up to 2.4, while the bin of $|Y|>2.4$ is not used in this work due to its low statistic. 
The $A_{FB}$ results with $M>320$ GeV are also excluded, due to their low sensitivity and large uncertainties from $\Delta(M)$.
The fitted $\effstw$ values, as given in Table~\ref{tab:CMSstw}, are statistically consistent with the value of $0.23101 \pm 0.00053$
measured by the CMS Collaboration using the same data~\cite{CMS8TeVStw}.
\begin{table}[hbt]
\begin{footnotesize}
\begin{center}
\begin{tabular}{l|c|c|c|c}
\hline \hline
$|Y|$ bins & [0, 1.0] & [1.0, 1.25] & [1.25, 1.5] & [1.5, 2.4] \\
\hline
Fitted & 0.2336 & 0.2323 & 0.2300 &  0.2313 \\
 $\effstw$ & $\pm$ 0.0017 & $\pm$ 0.0016 & $\pm$ 0.0016 & $\pm$ 0.0006 \\  
\hline \hline
\end{tabular}
\caption{\small Fitted values and uncertainties of $\effstw$ from the CMS 8TeV $A_{FB}(M)$ measurement. 
The uncertainty includes the fitting error derived from experimental uncertainty, and the theoretical 
error arising from $\Delta(M)$ estimated by using CT18 error sets.}
\label{tab:CMSstw}
\end{center}
\end{footnotesize}
\end{table}
The observed structure parameters 
$P^u_0$ and $P^d_0$, as a function of $|Y|$, are shown in Figure~\ref{fig:CMSP0}, compared to the {\sc ResBos} 
predictions with CT18, MSHT20 
and NNPDF3.1~\cite{CT18, MSHT20, NNPDF31} PDFs.
In all $|Y|$ bins, the  
observed $P^u_0$ values are smaller than the theory predictions, while the $P^d_0$ values are larger than the expectations. 
The deviation implies that there might be more significant contribution from the down-type 
quark subprocesses with respect to the theory prediction of current PDF sets. 
Such results reflect the behavior of the $A_{FB}(M)$ distributions reported by CMS.
Due to the difference between the $Z$ boson couplings to 
the down-type and up-type quarks, the 
magnitude of $A^d_{FB}$ around the $Z$ pole is smaller than that of $A^u_{FB}$.
Consequently, if the measured $A_{FB}$ values around $M_Z$ are closer to zero than expectation, 
it could naturally imply a higher weight of $A^d_{FB}$ in the data. 
This feature can be clearly seen through the CMS reported $A_{FB}$ 
around the $Z$ pole , as depicted in Figure~\ref{fig:CMSAFBvsPDF}.

In principle, as shown in Eq.~\eqref{eq:LHCAFBfactorization}, $P^u_0$ and $P^d_0$ 
contain various information.  
Their values are governed by the light quark ($u$ and $d$) PDFs at both 
$x_L$ and $x_S$ regions; The $s$, $c$ and $b$ quark contributions, which appear 
in the denominators in Eq.~\eqref{eq:LHCAFBfactorization}, 
can also change 
the observed $P_0$ values; 
It might even be complicated by taking the difference between $q$ and $\bar{q}$ densities 
for $s$, $c$ and $b$ quarks into account.

However, $\alpha_s$, $\alpha_c$ and $\alpha_b$ are not as large as 
$\alpha_u$ and $\alpha_d$, thus not dominating the $Z$ boson production. 
Contribution from the difference between $q$ and $\bar{q}$ densities for 
$q=s$, $c$ and $b$ is even smaller. Up to the next-to-leading order (NLO) in QCD, 
we have $c(x)=\bar{c}(x)$ and $b(x)=\bar{b}(x)$ at the $Q_0$ scale around 1 GeV, from 
where the PDFs are evolved to higher energy scales, so that $1-2D=0$. 
At NNLO, a non-vanishing $1-2D$ can be generated, but expected to be negligible due to the 
suppression of the strong coupling strength. For the $s$ quark, $s(x)\ne \bar{s}(x)$ is already 
allowed at the leading order (LO) in the global analysis of both MSHT20 and NNPDF4.0, 
but the difference between $s(x)$ and $\bar{s}(x)$ in the relevant $x$ region of this work 
does not induce noticeable contribution to $P_0$.
Therefore, the leading sensitivity of $P^u_0$ and $P^d_0$ as experimental observables is 
on the $u$ and $d$ quark PDFs. 
As discussed in the Introduction, recent global analyses yield a stronger $d$ quark 
 PDF after including the measurements of the single inclusive $W$ and $Z$ boson productions 
 (without $A_{FB}$) at the LHC.
Based on the above discussions, the results from the CMS $A_{FB}$ measurement, to a certain degree, 
support the conclusion from the recent global analysis, which is one possible explanation on the deviation 
between the measured $P_0$ values and the theory predictions.

Nevertheless, the measured structure parameters can now be used as standalone data constraints in the PDF global analysis, 
In Table~\ref{tab:CMSP0}, we list the $P^u_0$ and $P^d_0$ values extracted 
from the CMS 8 TeV $A_{FB}(M)$ data.
\begin{table}[hbt]
\begin{footnotesize}
\begin{center}
\begin{tabular}{l|c|c|c}
\hline \hline
 $|Y|$ bins & $P^u_0\pm(\text{exp.})$ & $P^d_0\pm(\text{exp.})$ & correlation \\
   & $\pm(\Delta)$ & $\pm(\Delta)$ & \\
\hline
 [0, 1.0] & $0.1118 \pm 0.0081$ & $0.0551\pm 0.0118$ & -0.92  \\
  & $\pm0.0030$ & $\pm 0.0039$ & \\
\hline
 [1.0, 1.25] & $0.2644\pm0.0176$ & $0.1116\pm 0.0247$ & -0.93  \\
  & $\pm0.0048$ & $\pm0.0073$ & \\
\hline
 [1.25, 1.5] & $0.3350\pm 0.0193$ & $0.1282\pm 0.0273$ & -0.93  \\
  & $\pm0.0053$ & $\pm0.0083$ & \\
\hline
 [1.5, 2.4] & $0.4681\pm 0.0155$ & $0.1955\pm 0.0193$ &  -0.92  \\
   & $\pm0.0069$ & $\pm0.0105$ & \\
\hline \hline
\end{tabular}
\caption{\small Fitted values and uncertainties of $P^u_0$ and $P^d_0$ from the CMS $A_{FB}(M)$ measurement. 
The first uncertainties in the breakdown are extrapolated from the experimental uncertainties on the $A_{FB}(M)$, 
with the bin-by-bin correlation on 
systematics taken into account. The second uncertainties in the breakdown correspond to the 
theoretical errors arising from $\Delta(M)$ estimated by using the CT18 error sets.}
\label{tab:CMSP0}
\end{center}
\end{footnotesize}
\end{table}
~\\

\section{Structure parameters from the Tevatron data}

In this section, we extract the structure parameters from the $A_{FB}(M)$ spectrum measured in $p\bar{p}\rightarrow 
Z/\gamma^* \rightarrow e^+e^-$ events at $\sqrt{s}=1.96$ TeV by the D0 collaboration ~\cite{D0AFB5fb}.  
Unlike the LHC, 
the hard processes in Tevatron $p\bar{p}$ collision are dominated by the valence $u$ and $d$-quarks. Besides, due to 
a relatively low beam energy, even the smaller momentum fraction $x_S$ at the Tevatron is around $10^{-2}$. As a result, 
the Tevatron data could provide a direct constraint especially in the $x$ 
region above 0.01, on the valence $u$ and $d$ quark PDFs.

At the Tevatron, $A_{FB}$ can be measured in the Collins-Soper frame, of which 
the $z$-axis is defined according to the directions of the proton and antiproton beams.
The factorization of the $A_{FB}(M)$ in $p\bar{p}\rightarrow Z/\gamma^*\rightarrow \ell^+\ell^-$ events shares exactly 
the same form in Eq.~\eqref{eq:LHCAFBfactorization}, with a different definition on the dilution factors. 
At the Tevatron, the dilution factor is defined as the probability of having a quark from the antiproton beam and an 
antiquark from the proton beam, namely both partons are governed by the PDFs of antiquarks in proton, 
and can be written as:
\begin{eqnarray}
 D_q(x_1, x_2) = \frac{ \bar{q}(x_1)\bar{q}(x_2)}{q(x_1)q(x_2)+\bar{q}(x_1)\bar{q}(x_2)}
\end{eqnarray}
where no requirement of $x_1>x_2$ or $x_1<x_2$ is needed.
Nonetheless, 
the dilution factors at the Tevatron are small anyway. They are in general lower than $10\%$, while at the LHC they 
can be as large as $40\%$ in low rapidity region. 

\begin{figure}[!hbt]
\begin{center}
\epsfig{scale=0.4, file=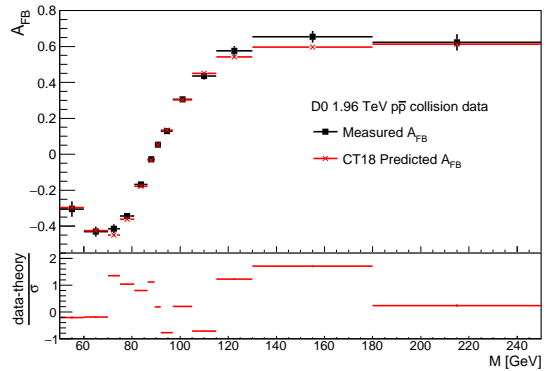}
\caption{\small The spectrum of $A_{FB}(M)$ measured using the Tevatron data, and 
the corresponding uncertainties.
The bottom panel is the difference between the D0 measured 
$A_{FB}$ and the {\sc ResBos}+CT18 predicted ones. The uncertainties on the theoretical predictions 
correspond to the $68\%$ C.L. PDF uncertainties}
\label{fig:D0AFBvsPDF}
\end{center}
\end{figure}

The D0 collaboration provided their $A_{FB}(M)$ results in a $Y$-integrated phase space, 
in a mass range up to 1 TeV. 
In this paper, we use the data in the mass window of [50, 250] GeV to extract the structure parameters. 
Higher mass region is excluded due to their low statistic and large uncertainty from $\Delta(M)$. 
In Figure~\ref{fig:D0AFBvsPDF}, we compare the D0 $A_{FB}(M)$ data and
the {\sc ResBos} prediction with CT18NNLO, as a function of $M$. 

The comparison shows the same tendency as the CMS data, 
that the observed asymmetry $A_{FB}$ at the Tevatron has smaller absolute 
values around $Z$ pole than predictions. 
The extracted values of the $P^u_0$ and $P^d_0$, 
together with their uncertainties, are compared to the {\sc ResBos} predictions 
using various PDFs in Table~\ref{tab:D0P0}. The correlation between the 
uncertainties of the extracted $P^u_0$ and $P^d_0$ is $-0.95$.
As expected, the observed structure parameters 
indicate more significant contribution from the $d$ quarks.
\begin{table}[hbt]
\begin{footnotesize}
\begin{center}
\begin{tabular}{l|c|c}
\hline \hline
& $P^u_0$ & $P^d_0$ \\
\hline
D0 data & $0.6395\pm0.0356(\text{exp.})$ &$0.2706\pm0.0662(\text{exp.})$ \\
 & $\pm0.0059(\Delta)$ & $\pm 0.0061(\Delta)$ \\
\hline
CT18 & $0.6994\pm0.0089$ &$0.1733\pm0.0062$ \\
\hline
MSHT20 & $0.6887\pm 0.0066$ & $0.1658\pm 0.0075$ \\
\hline
NNPDF3.1 & $0.6919\pm 0.0054$ & $0.1703\pm0.0055$ \\
\hline \hline
\end{tabular}
\caption{\small Predictions on $P^u_0$ and $P^d_0$ in $p\bar{p}$ collisions from CT18, MSHT20 and NNPDF3.1, 
compared to the extracted values from the D0 $A_{FB}(M)$. 
The first uncertainty labeled with exp. on the extracted $P^u_0$ and $P^d_0$ 
corresponds to the experimental uncertainty, while the second on labeled with $\Delta$ comes from 
the theoretical error of $\Delta(M)$ estimated by using the CT18 error sets.
Uncertainties on the predictions correspond to 
the $68\%$ C.L. PDF uncertainties.}
\label{tab:D0P0}
\end{center}
\end{footnotesize}
\end{table}
In fact, $P^u_0$ and $P^d_0$ reflect the relative strength of the $u\bar{u}$ and $d\bar{d}$ subprocesses in 
Drell-Yan productions, rather than their absolute contributions. Accordingly, when the observed $d$ quark 
contributions are enhanced, the $u$ quark ones are expected to be suppressed. These negative 
correlations have been demonstrated in both the observations of $P^u_0$ and $P^d_0$ form the CMS and D0 data.

On the other hand, the fitted $\effstw$ gives $0.2318 \pm 0.0014$, which is consistent with the value of $0.2309 \pm 0.0010$ 
extracted by the D0 Collaboration~\cite{D0AFB5fb}, using the same data with conventional method. In fact, as concluded in 
Ref.~\cite{AFBshape}, the PDFs change the $A_{FB}(M)$ distribution on its shape as a rotation around the $Z$ pole, 
while $\effstw$ 
governs $A_{FB}(M)$ more on its average value. Therefore, 
both the results of the Figure~\ref{fig:CMSAFBvsPDF} and Figure~\ref{fig:D0AFBvsPDF} 
call for a change in their corresponding $P^u_0$ and $P^d_0$ values, rather than 
the $\effstw$ value. That is indeed what we have found.

~\\

\section{Conclusion}

In this paper, we present the first application of the factorization 
formalism of the forward-backward asymmetry $A_{FB}$ observable at hadron 
colliders, and determine the proton structure parameters $P^u_0$ and $P^d_0$ 
by fitting to the $A_{FB}$ distributions measured by CMS and D0. 
The values of $P^u_0$ and $P^d_0$,  
determined from both the CMS and D0 experiments, are standalone observables and can be used as experimental inputs 
in the PDF global analysis. Though the observed 
structure parameters 
are still statistically limited, the CMS and D0 data coincidently hint at an indication that the 
down-type quark contribution might be higher than the predictions of current PDFs.
 
We would like to point out that: 1) The indication by now simply comes from the direct comparison between the extracted 
values of $P^u_0$ and $P^d_0$ and their theoretical predictions 
based on the factorization formalism presented in Ref.~\cite{AFBFactorization}. 
To understand the impact of the structure parameter 
measurements, the numerical results of this work should be introduced into a complete PDF global analysis; 
2) To confirm 
the deviation of observed $P^u_0$ and $P^d_0$, larger data sample should be used at both hadron colliders. 
For the LHC, the 130 fb$^{-1}$ data at 13 TeV 
has already been collected during its Run 2 period, and more data will be collected in the 
future. For the Tevatron, the $A_{FB}(M)$ distribution used in this work corresponds to only half 
of the D0 data with one single channel of the dielectron final state.
It could be several times more events if the full dataset collected by both the D0 and CDF detector can be used, 
with both dielectron and dimuon final states included.

\section{Acknowledgements}
This work was supported by the National Natural Science Foundation of China under Grant No. 11721505, 11875245, 
12061141005 and 12105275, and supported by the ``USTC Research Funds of the Double First-Class Initiative''. 
This work was also supported by the U. S. National Science Foundation under Grant No. PHY-2013791. 
C.-P. Yuan is also grateful for the support from the Wu-Ki Tung endowed chair in particle physics.

\end{document}